\documentclass[draft]{aipproc}
\newcommand{\be}{\begin{equation}}
\newcommand{\ee}{\end{equation}}
\newcommand{\bold}{\textbf}
\usepackage{graphicx}
\layoutstyle{6x9}

\makeatletter
   \def\@oddfoot{\hfil \thepage}

\begin{document}

\title{Big Bang Nucleosynthesis and the Missing Hydrogen Mass in the
Universe\thanks{Talk given at CIPANP2003.  To be published by the
American Institute of Physics in the Proceedings of the Conference
on the Intersections of Particle and Nuclear Physics (CIPANP2003)
held at Grand Hyatt Hotel -- New York City, May 19-24, 2003.}}

\author{D. C. Choudhury}{address={Department of Physics, Polytechnic University,
Brooklyn, New York 11201, USA, \\ email: dchoudhu@duke.poly.edu}}
\author{David W. Kraft}{address={Dana Hall, University of Bridgeport, Bridgeport, CT
06601, USA}}

\begin{abstract}
    It is proposed that when the era of the big-bang nucleosynthesis
    ended, almost all of the 75 percent of the observed total baryonic
    matter remained in the form of hydrogen and continued
    to exist in the form of protons and electrons.  They are present
    today as baryonic dark matter in the form of intergalactic hydrogen
    plasma.  To test our hypothesis we have investigated the effects of
    Thomson scattering by free electrons on the reported dimming of Type
    Ia supernovae.  The quantitative results of our calculation suggest
    that the dimming of these supernovae, which are dimmer than expected
    and hence more distant than predicted by Hubble expansion, is a result
    of Thomson scattering without cosmic acceleration.
\end{abstract}

\maketitle

Recent observations [1-2] of Type Ia supernovae (SNe Ia) appear to
suggest that the universe is accelerating.  The basic idea
proposed to account for the acceleration is dark energy or
quintessence [3-6].  However, as yet there is no direct
confirmation of their existence or exact nature.  The present work
examines whether the question of acceleration can be resolved
within the limits of established laws of physics.  Consequently,
we have investigated the effects of Thomson scattering of photons
by free electrons present in the form of H-plasma while
propagating from the supernova source to the point of observation.
This particular scattering process has been chosen for the
following reasons: the scattering cross-section is pure elastic in
nonrelativistic region, the total cross-section is a universal
constant and it is independent of the incident frequency [7].
Hence characteristics of the atomic spectra, which are relevant in
the present work, remain unchanged. The calculation is performed
within the framework of Friedmann-Robertson-Walker (FRW) cosmology
[8] for the special case of a flat universe, consistent with the
recent cosmic microwave background anisotropy measurements [9,10]
indicating a spatially flat, critical density universe with
$\Omega=1$ and with the inflationary model [11-13] of cosmology.
Hence in the present investigation we consider a universe, for
large-scale structure, consisting only of matter and negligible
radiation without cosmological constant in accord with all popular
cosmological models prior to the late 1990s.  Therefore in terms
of standard notation, \be \Omega  = \frac{{\rho (z)}}{{\rho _c
(z)}} = 1\ee and \be  \rho _c (z) = \frac{{3H^2 (z)}}{{8\pi G}} =
\frac{{3H_0^2 }}{{8\pi G}}(1 + z)^3\ee where $\rho$ is the total
mass density including radiation, $\rho_c$ is the critical
density, $G$ is the gravitational constant, $H$ is the Hubble
constant and in which we have incorporated the flat universe
relation between $H(z)$ and the current Hubble constant $H_0$.

In light of the above considerations, the total matter density of
universe consists of ordinary baryonic matter, baryonic dark
matter and nonbaryonic dark matter; thus we have  $\Omega =
\Omega_{bm} + \Omega_{bdm} + \Omega_{nbdm}$ where each term is in
units of $\rho_c$, and where the subscripts refer, respectively,
to the three aforementioned components.  The observed mass density
$\Omega_m \approx 0.30 \pm 0.10$ as estimated from its
gravitational pull on visible matter and is assumed to consist of
ordinary baryonic matter and baryonic dark matter instead of
exotic dark matter [14]; thus  $\Omega_m = \Omega_{bm} +
\Omega_{bdm}$. Big bang nucleosynthesis and synthesis of heavy
elements in stars have been extensively investigated for more than
half a century. The quantitative results of atomic abundances of
various groups of elements [15-16] by mass-fraction of the total
are hydrogen  $\cong 0.75$ and rest of the elements $\cong 0.25$.
From the above value of $\Omega_m$ we obtain the magnitude of
$\Omega_{bm}$ which includes all the elements formed during big
bang nucleosynthesis, and of $\Omega_{bdm}$ which remained as free
hydrogen (in the form of mostly protons and equal number of
electrons, neutrons having decayed into protons and electrons)
within half an hour after the era of big bang nucleosynthesis
ended.  These magnitudes are $0.05 \leq \Omega_{bm} \leq 0.10$ and
$0.15 \leq \Omega_{bdm} \leq 0.30$. From this analysis we conclude
that when the era of the big bang nucleosynthesis ended, the
universe continued expanding for several thousand years or so
until the temperature dropped low enough to form neutral atoms. We
believe that it is in this time interval that most of the free
protons and electrons (contained in $\Omega_{bdm}$) escaped into
cosmic space and are most likely present in the form of an
intergalactic hydrogen plasma [17]. They are dark because they
cannot emit light.
\begin{table}[t]
    \begin{tabular}{ccccccc} \hline
             & Distance & Distance            & \multicolumn{3}{c}{\underline{With Thomson scattering correction}} & Theoretical \\
    Redshift & modulus  & $R_{\mathrm{obs}}$  & $\Omega_{bdm}=0.15$ & $\Omega_{bdm}=0.23$ & $\Omega_{bdm}=0.30$    & distance $D_L$ \\
    $z$      & ($m-M$)  & (Mpc)               & (Mpc)               & (Mpc)               & (Mpc)                  & (Mpc) \\ \hline
    0.0043 & 31.72 & 22 & 22 & 22 & 22 & 20 \\
    0.0077 & 32.81 & 36 & 36 & 36 & 36 & 36 \\
    0.025 & 35.35 & 117 & 117 & 117 & 117 & 115 \\
    0.052 & 36.72 & 221 & 219 & 218 & 217 & 241 \\
    0.053 & 37.12 & 265 & 263 & 262 & 261 & 248 \\
    0.068 & 37.58 & 328 & 324 & 322 & 320 & 318 \\
    0.090 & 38.51 & 504 & 495 & 491 & 487 & 424 \\
    0.17 & 39.95  & 977 & 945 & 929 & 915 & 825 \\
    0.30 & 41.38 & 1888 & 1775 & 1718 & 1670 & 1474 \\
    0.38 & 41.63 & 2118 & 1956 & 1874 & 1807 & 1893 \\
    0.43 & 42.15 & 2692 & 2454 & 2336 & 2238 & 2160 \\
    0.44 & 41.95 & 2455 & 2234 & 2296 & 2035 & 2214 \\
    0.48 & 42.39 & 3006 & 2783 & 2560 & 2438 & 2430 \\
    0.50 & 42.40 & 3020 & 2707 & 2553 & 2428 & 2539 \\
    0.57 & 42.76 & 3565 & 3138 & 2931 & 2766 & 2924 \\
    0.62 & 42.98 & 3945 & 3428 & 3179 & 2982 & 3203 \\
    0.83 & 43.67 & 5420 & 4447 & 4000 & 3656 & 4400 \\
    0.97 & 44.39 & 7551 & 5937 & 5221 & 4681 & 5225 \\ \hline
    \end{tabular}
    \caption{Distance moduli are those reported in ref.~[1].  Data at $z =
    0.43$ and 0.48 are the mean values of two observations at each of these
    redshifts.  Distances in columns~3-6 are computed from eq.~(8).  Column~3
    lists distances computed from observed values of $m-M$ and the values in
    columns~4-6 result from distance moduli corrected for Thomson scattering;
    these three columns correspond to the extreme values of the hydrogen mass
    fraction $\Omega_{bdm}$ cited above and to their midpoint,
    $\Omega_{bdm} \simeq 0.23$.  Values of the theoretical luminosity
    distance in the last column are computed
    from eq.~(9).  The best agreement with the theoretical values in column~7
    are those listed for $\Omega_{bdm} \simeq 0.23$ in column~5.}
\end{table}

The intensity of the radiation lost to Thomson scattering depends
critically on the density of free electrons in the path from the
source to the observer.  We take the free electron number density
at redshift $z$ as \be n(z)=\frac{\Omega_{bdm}\rho_c(z)}{m_h}\ee
where $\Omega_{bdm}\rho_c$ is the hydrogen mass density and $m_h$
is the hydrogen mass.  We consider radiation emitted by a Type Ia
supernova at redshift $z_S$ and received by an infinitesimal
volume element of length $cdt$ at redshift $z$.  The fractional
reduction of intensity owing to Thomson scattering by free
electrons within this volume is \be \frac{{ - dI}}{{I(z)}} =
\sigma _T n(z)cdt\ee where $I(z)$ is the incident intensity and
$\sigma_T$ is the total Thomson scattering cross-section by a free
electron. The total attenuation by scattering is the integral of
eq.~(7) over the entire path taken by the light from the
supernova.  We perform the integration by identifying $dt$ with
$dT_H(z)$ where $T_H(z)$ is the Hubble time $H^{-1}(z)$.  Thus \be
dt=dT_H(z)=-\frac{3}{2}H_0^{-1}(1+z)^{-5/2}dz.\ee Combining
eqs.~(2)-(5) then yields \be
\frac{dI}{I}=\sigma_T\Omega_{bdm}\frac{9cH_0}{16\pi
Gm_h}(1+z)^{1/2}dz.\ee  Integrating over a path from the source to
the observer at $z=0$ yields \be I(0)=I_S\exp
\left\{-\sigma_T\Omega_{bdm}\frac{3cH_0}{8\pi
Gm_h}[(1+z)^{3/2}-1]\right\}\ee where $I_S$ is the light intensity
at the source and in which the exponential factor represents the
loss of intensity by Thomson scattering.

Observations of redshift and distance modulus $m-M$ for Type Ia
supernovae reported by Riess et al. [1] are listed in Table 1 in
columns 1 and 2.  Column 3 contains the distance computed for each
distance modulus according to [18] \be
R_\textrm{obs}=10^{(m-M-25)/5}\ee where $R_{\textrm{obs}}$ is the
distance in Mpc. Corrections for Thomson scattering are effected
by multiplying the observed values of $m-M$ by the ratio
$I(0)/I_S$ from eq.~(7) and columns 4-6 contain the corrected
distances as computed from eq.~(8); these three columns correspond
to the extreme values of the hydrogen mass fraction $\Omega_{bdm}$
listed above and to their midpoint, $\Omega_{bdm} \cong 0.23$. The
calculations employ $H_0 = 65$~km-s$^{-1}$-Mpc$^{-1}$.  The last
column lists values of the theoretical luminosity distance $D_L$
which, for a flat universe with deceleration parameter $q_0 =
1/2$, is given by [8] \be D_L=\frac{2c}{H_0}(1+z-\sqrt{1+z}).\ee

\emph{Conclusion.}  The effects of Thomson scattering on enhanced
dimming of SNe Ia presented in Table 1 suggests that (i) the
recently observed supernovae data can be understood without dark
energy; (ii) the amounts of baryonic ordinary matter, baryonic
dark matter and nonbaryonic dark matter are in the ranges of
5-10\%, 15-30\% and 60-80\%, respectively; and (iii) the total
matter density $\Omega$ consists of 5-10\% baryonic ordinary
matter and 90-95\% total dark matter, consistent with the
understanding of most cosmologists prior to late 1990s.  Further
details of the present investigation will be published elsewhere.

\begin{center}\textbf{REFERENCES}\end{center} \vspace{0.5em}

1.  Riess, A.G. et al, \emph{Astron. J.} \bold{116}, 1009-1038
(1998).

2. Perlmutter, et al, \emph{Astrophys. J.} \bold{517}, 565-586
(1999).

3.  Wang, L. and Steinhardt, P. J., \emph{Astrophys. J.}
\bold{508}, 483-490 (1998).

4. Wang, L., Caldwell, R. R., Ostriker, J. P. and Steinhardt, P.
J., \emph{Astrophys. J.}

\hspace{1em} \bold{530}, 17-35 (2000).

5. Perlmutter, S., Turner, M. S. and White, M., \emph{Phys. Rev.
Lett.} \bold{83}, 670-673

\hspace{1em} (1999).

6. Peebles, P. J. E. and  Ratra, B., \emph{Rev. Mod. Phys.}
\bold{75}, 559-606 (2003).

7. Heitler, W., \emph{The Quantum Theory of Radiation}, 3rd
ed.,(Dover, New York,

\hspace{1em} 1984), pp. 34-35.

8.  Kolb, E. W. and Turner, M. S., \emph{The Early Universe},
paperback ed., (Addison-

\hspace{1em} Wesley, Massachusetts, 1994), pp.~47-86.

9.  de Bernardis, P. et al., \emph{Nature} \bold{404}, 955-959
(2000).

10. Netterfield, C. B. et al., \emph{Astrophys. J.} \bold{571},
604-614 (2002).

11. Guth, A. H., \emph{Phys. Rev.} \bold{D23}, 347-356 (1981).

12. Linde, A. D., \emph{Phys. Lett.} \bold{B108}, 389-393 (1982).

13. Albrecht, A. and Steinhardt, P. J., \emph{Phys. Rev. Lett.}
\bold{48}, 1220-1223 (1982).

14. Bahcall, N. A., Ostriker, J. P., Perlmutter, S. and
Steinhardt, P. J.,

\hspace{1em} \emph{Science} \bold{284}, 1-16 (1999).

15. Burbidge, E. M., Burbidge, G. B., Fowler, W. A. and Hoyle, F.,
\emph{Rev. Mod.}

\hspace{1em} \emph{Phys.} \bold{29}, 547-650 (1957).

16. Boesgaard, A. M. and Steigman, G., \emph{Ann. Rev. Astron.
Astrophys.} \bold{23},

\hspace{1em} 319-378 (1985).

17. Weinberg, S., \emph{Gravitation and Cosmology} (Wiley, New
York, 1972), p. 500.

18. Peebles, P. J. E., \emph{Principles of Physical Cosmology},
(Princeton Univ. Press,

\hspace{1em} Princeton, 1993), p.21.

\end{document}